\documentclass[sigconf]{acmart}

\usepackage{algorithm}
\usepackage{algpseudocode}
\usepackage{multirow}
\usepackage{xcolor}

\AtBeginDocument{%
  }

\copyrightyear{2026}
\acmYear{2026}

\setcopyright{cc}
\setcctype{by}

\acmConference[CIKM '26]
  {Proceedings of the 35th ACM International Conference on Information and Knowledge Management}
  {November 7--11, 2026}
  {Rome, Italy.}

\acmBooktitle
  {Proceedings of the 35th ACM International Conference on Information and Knowledge Management (CIKM '26), November 7--11, 2026, Rome, Italy}

\acmISBN{979-8-4007-2539-5/2026/11}

\acmDOI{10.1145/3799682.3840121}

\begin{document}

\title[SAM-D2Q: Multimodal Doc2Query for E-commerce Search]
{SAM-D2Q: Aligning Multimodal Doc2Query with Search Demand and Conversion for E-commerce}

\author{Hui Zhou}
\email{shouta.zh@alibaba-inc.com}
\authornote{Corresponding author.}
\affiliation{%
  \institution{Alibaba International Digital Commerce Group}
  \city{Hangzhou}
  \state{Zhejiang}
  \country{China}
}

\author{Jian Hui Ji}
\email{jianhui.jjh@alibaba-inc.com}
\affiliation{%
  \institution{Alibaba International Digital Commerce Group}
  \city{Hangzhou}
  \state{Zhejiang}
  \country{China}
}

\author{Lei Ma}
\email{ml373836@alibaba-inc.com}
\affiliation{%
  \institution{Alibaba International Digital Commerce Group}
  \city{Hangzhou}
  \state{Zhejiang}
  \country{China}
}

\author{Rong Xiao}
\email{xiaorong.xr@taobao.com}
\affiliation{%
  \institution{Alibaba International Digital Commerce Group}
  \city{Hangzhou}
  \state{Zhejiang}
  \country{China}
}

\author{Xiaoyi Zeng}
\email{yuanhan@taobao.com}
\affiliation{%
  \institution{Alibaba International Digital Commerce Group}
  \city{Hangzhou}
  \state{Zhejiang}
  \country{China}
}

\renewcommand{\shortauthors}{Hui Zhou, Jian Hui Ji, Lei Ma, Rong Xiao, \& Xiaoyi Zeng}

\begin{abstract}
E-commerce search often suffers from vocabulary mismatch between user queries and merchant-authored product titles, since short titles cannot fully cover diverse user expressions or visual product attributes. Although Doc2Query alleviates this issue by generating pseudo-queries for document expansion, traditional methods are text-only and not optimized for e-commerce business objectives. As a result, they may produce semantically plausible but commercially ineffective expansions and miss key attributes present in product images. To this end, we propose \textbf{E-commerce Search-Aligned Multimodal Doc2Query} (\textbf{SAM-D2Q}), a business-aligned multimodal document expansion framework for e-commerce search under Boolean retrieval constraints. SAM-D2Q consists of three stages: (1) task-adapted multimodal supervised fine-tuning to enhance vision-language understanding of product titles, images, and user queries; (2) multimodal data augmentation to improve perception of key visual attributes and expansion coverage; and (3) reinforcement-learning-based preference alignment toward search business objectives, encouraging the model to generate pseudo-queries that better match user intent and commercial value. Offline experiments show that SAM-D2Q substantially improves retrieval performance over traditional Doc2Query methods. Deployed in the AliExpress production search system, SAM-D2Q improves online business metrics, increasing GMV by \textbf{+3.38\%} and Pay Count by \textbf{+2.27\%}.
\end{abstract}

\ccsdesc[500]{Information systems~Language models}

\keywords{E-commerce Search, Sparse Retrieval, Doc2Query, Vision-Language Models, Preference Alignment}

\maketitle

\section{Introduction}

With the rapid growth of e-commerce, online shopping has become an integral part of daily life for billions of users worldwide. In this digital ecosystem, search engines serve as the key gateway connecting user intent with massive product inventories. A typical e-commerce search pipeline consists of query understanding, retrieval/recall, and ranking. Although neural retrieval methods such as dense retrieval and generative retrieval have made significant progress, keyword-based retrieval built on inverted indexes remains the backbone of large-scale e-commerce search systems due to its low latency, high interpretability, robustness for cold-start products, and reliability in exact matching scenarios such as brands and models.

However, keyword-based retrieval still suffers from the vocabulary mismatch problem, where user queries and product descriptions may express the same intent using different terms. To alleviate this issue, document expansion methods have been widely studied. In particular, Doc2Query~\cite{gospodinov2023doc2query,nogueira2019doc2query,nogueira2019document}, which enriches document representations by generating potential user queries, has achieved strong performance in general web search. Nevertheless, directly applying traditional Doc2Query methods to e-commerce search faces two major challenges.

First, existing Doc2Query methods mainly optimize semantic relevance based on textual data, while overlooking critical business signals such as search Page Views (PV) and Conversion Rate (CVR). As a result, the generated queries may be semantically related but have limited commercial value. For example, in the ``dress'' category, high-performing queries often contain concrete style attributes such as ``sexy dress'' or ``casual dress'', which typically correspond to higher search demand or conversion potential. In contrast, a generic Doc2Query model may generate low-value variants such as ``chic dress'' or simply reorganize existing title words, introducing index redundancy without clear business gain.

Second, e-commerce products are inherently multimodal, while product titles are often short and incomplete due to length constraints. Many key attributes, such as color, style, pattern, and material, are mainly expressed in product images rather than text. For long-tail products with sparse titles, such as ``man shirt 2025'', text-only expansion methods may fail to capture visual details such as sleeve length or specific colors, causing relevant products to be missed during retrieval.

To address these challenges, we propose \textbf{SAM-D2Q}, a \textbf{Search-Aligned Multimodal Doc2Query} framework for e-commerce search. Instead of focusing only on semantic enrichment, SAM-D2Q aligns query generation with search business objectives while incorporating multimodal product information. By integrating visual signals, multimodal data synthesis, and business-aligned optimization, SAM-D2Q generates expanded queries that are both grounded in product content and valuable for downstream commercial performance.

SAM-D2Q consists of three stages:
(1) information-gain constrained multimodal SFT for coarse semantic
alignment;
(2) hybrid multimodal SFT with CPV-guided counterfactual augmentation
for visual grounding; and
(3) GRPO-based preference alignment toward search business objectives.

The novelty of SAM-D2Q lies in the industrial formulation and deployment
of multimodal Doc2Query under Boolean retrieval constraints, where
expansion terms must provide recall gain, visual grounding, and business
utility. Deployed in AliExpress Search, SAM-D2Q brings a 3.38\% GMV lift
and a 2.27\% Pay Count increase in online A/B testing.

The main contributions of this work are summarized as follows:
\begin{itemize}
    \item \textbf{Industrial multimodal Doc2Query formulation.}
    We reformulate Doc2Query for Boolean e-commerce retrieval, where
    expansion terms must provide information gain rather than TF
    re-weighting, and align generation with recall, visual grounding,
    and business utility.

    \item \textbf{Counterfactual visual grounding.}
    We convert text-covered logs into multimodal counterfactual samples
    by masking visual attribute terms from product text while preserving
    product images, forcing the model to recover visually grounded
    expansion targets.

    \item \textbf{Large-scale deployment.}
    We deploy SAM-D2Q in AliExpress Search via offline indexing without
    online model inference, and validate it through offline experiments
    and online A/B testing.
\end{itemize}

\section{Related Work}

\subsection{Document Expansion}
Document expansion aims to alleviate vocabulary mismatch by enriching
documents with additional terms before indexing. Given a document
$d_i$, Doc2Query-style methods generate synthetic queries
$Q_i=\{q_{i1},\ldots,q_{im}\}$ and append them to the original document
for retrieval. Early Doc2Query~\cite{nogueira2019document} uses a
Seq2Seq Transformer to generate queries, while DocT5Query
~\cite{nogueira2019doc2query} further improves effectiveness with a
larger T5 model. Recent studies also explore LLM-based query generation,
but their advantages over strong T5-based baselines remain mixed
~\cite{weller2024generative}. Since generative expansion may introduce
irrelevant or hallucinated terms, Doc2Query-{}-~\cite{gospodinov2023doc2query}
filters generated queries with a relevance model to reduce noise.

However, e-commerce search differs substantially from general web
retrieval. Industrial systems typically follow a recall-ranking cascade,
where the sparse retrieval stage prioritizes high recall under strict
latency constraints. Moreover, product titles are short and incomplete,
while many key attributes are expressed visually in product images.
Finally, e-commerce expansion should not only be semantically correct
but also commercially valuable, since query traffic and conversion rates
vary significantly across terms. These differences motivate our
business-aligned multimodal Doc2Query framework.

\subsection{Multimodal Data Augmentation}
Multimodal Large Language Models (MLLMs), such as LLaVA
~\cite{liu2023visual} and Qwen-VL~\cite{bai2023qwen,bai2025qwen3vltechnicalreport},
have shown strong vision-language understanding capabilities. However,
studies in Visual Question Answering (VQA) show multimodal models
may over-rely on language priors and ignore visual evidence
~\cite{agrawal2018don,goyal2017making,niu2021counterfactual}. For
example, if most answers to ``What color are the bananas?'' are
``yellow'', models may learn dataset shortcuts instead of grounding the
answer in the image.

To mitigate such biases, prior work constructs counterfactual or
augmented samples by modifying images, questions, or their associations.
MUTANT~\cite{gokhale2020mutant} creates semantically distinct samples
through input perturbation, CSS~\cite{chen2020counterfactual} masks
critical objects or words, and SimpleAug~\cite{kil2021discovering}
leverages semantic annotations for question pairing. Inspired by these
ideas, we construct CPV-guided counterfactual samples for e-commerce
products, masking visual attributes from text while preserving images,
so that the model must recover visually grounded attributes from product
images.

\subsection{Preference Alignment via Reinforcement Learning}
Supervised fine-tuning provides basic instruction-following ability, but
it does not necessarily align generation with downstream objectives.
RLHF, commonly implemented with PPO~\cite{schulman2017proximal,ouyang2022training},
is widely used to align LLMs with human preferences, but it is often
costly and unstable. Offline alternatives such as DPO~\cite{rafailov2023direct}
and IPO~\cite{azar2024general,ethayarajh2024kto} avoid explicit reward
modeling, but are limited by static preference data and weaker online
exploration.

Recent online RL methods such as Group Relative Policy Optimization
(GRPO)~\cite{shao2024deepseekmath} reduce training overhead by removing
the value network while retaining group-based exploration. Unlike
standard safety- or helpfulness-oriented alignment, we adapt GRPO to
optimize e-commerce query generation with semantic, commercial, and
visual rewards. This aligns document expansion with both retrieval
effectiveness and business utility.

\section{Preliminaries}
\label{sec:preliminaries}

E-commerce retrieval operates under distinct paradigms compared to general web search. 

The nature of product data and the reliance on Boolean matching necessitate a specialized approach to document expansion.

\subsection{The Necessity of Boolean-Aware Expansion}
Unlike general web search, where probabilistic models like BM25 are standard, industrial e-commerce search relies heavily on \textbf{Bool\-ean Matching} for its inverted index. This choice is mandated by three operational requirements:
\begin{enumerate}
    \item \textbf{Deterministic Matching:} To handle queries with strict intent slots (e.g., brand or model constraints), Boolean logic ensures precision by prohibiting the "fuzzy" retrieval typical of probabilistic models.
    \item \textbf{Data Conciseness:} Product titles act as structured keyword sets rather than natural language, rendering TF-based relevance weighting largely ineffective.
    \item \textbf{Latency Constraints:} Boolean retrieval offers superior computational efficiency at the scale of hundreds of millions of items, which is critical for maintaining real-time system performance.
\end{enumerate}

These factors imply that document expansion in this domain must not just "add keywords" as in Doc2Query, but must produce high-precision, business-aligned terms that strictly comply with the Boolean retrieval logic—a challenge that motivates our proposed SAM-D2Q framework.

\subsection{Implications for Document Expansion}
This paradigm shift fundamentally alters the objective of Doc2Query. 
In standard literature \cite{lin2022pretrained}, generated queries serve two purposes: (1) \textit{Expansion} (adding new terms) and (2) \textit{Re-weighting} (repeating existing terms to boost TF).
However, under the Boolean Matching constraint, \textit{re-weighting is
functionally redundant}—a term's presence is binary (0 or 1). Repeating
a token that already exists in the title provides \textit{zero marginal
utility} for recall.

Consequently, in our system context, the value of a generative expander is strictly defined by its ability to generate \textit{new terms}—latent keywords that are semantically relevant but lexically absent from the seller's description. This insight directly motivates our \textit{Information-Gain Constraint}
design in \textbf{Stage 1} (Section~\ref{sec:stage1}), where we use
non-covered query-item pairs for lexical expansion and reserve
text-covered pairs for counterfactual visual augmentation. This
Boolean-aware formulation distinguishes SAM-D2Q from standard
Doc2Query methods that are primarily designed for probabilistic
ranking-oriented retrieval.

\section{Methodology}
\label{sec:method}

Formally, we define the e-commerce document expansion task as learning
a conditional generation policy $\pi_\theta(Q \mid \mathcal{X})$,
parameterized by a Multimodal Large Language Model \textbf{(MLLM)}.
Here, $\mathcal{X} = \{\mathcal{T}, \mathcal{I}\}$ represents the
multimodal product context, consisting of textual metadata
$\mathcal{T}$ and visual content $\mathcal{I}$, while $Q$ denotes a
sequence of tokens representing a potential user search query.
Unlike traditional retrieval methods that rely on exact matching, our {LLM-based Doc2Query} approach leverages the extensive world knowledge and reasoning capabilities of the pre-trained MLLM to infer latent user intents that are semantically implied but lexically absent in the source document.

\begin{figure*}[t]  
    \centering
    \includegraphics[width=1\textwidth]{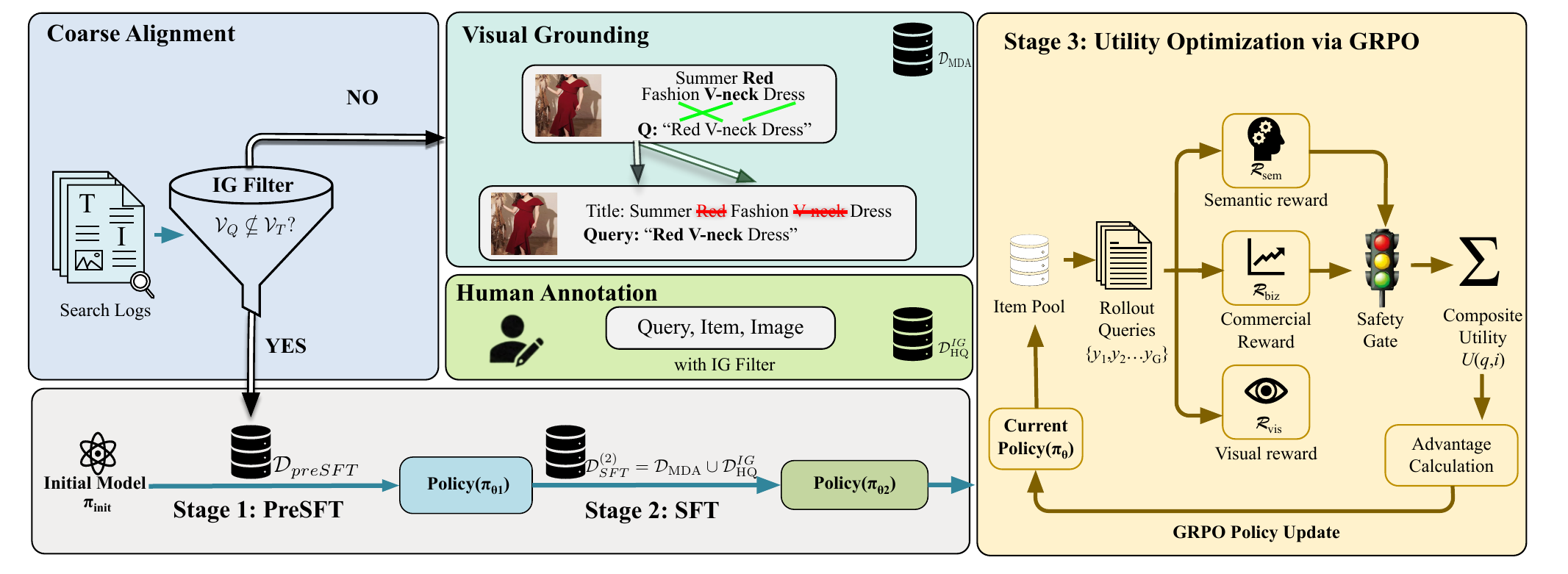} 
   
    \caption{\textbf{The overall pipeline of SAM-D2Q.}
SAM-D2Q contains three stages: 
(1) \textbf{Coarse Alignment}, where logs satisfying
$\mathcal{V}_Q \nsubseteq \mathcal{V}_T$ form
$\mathcal{D}_{\mathrm{preSFT}}$ and text-covered pairs are reserved for
augmentation;
(2) \textbf{Visual Grounding}, where visual attributes are masked from
product text to construct $\mathcal{D}_{\mathrm{MDA}}$, which is combined
with $\mathcal{D}_{\mathrm{HQ}}^{IG}$ for Stage-2 SFT;
and (3) \textbf{Utility Optimization}, where GRPO optimizes a gated
utility combining semantic, commercial, and visual rewards.}
    \Description{A framework diagram illustrating the SAM-D2Q pipeline. 
    The pipeline is divided into three stages. Stage 1 on the left splits 
    search logs by the information-gain constraint: pairs whose query tokens 
    are not fully covered by product text are used for pre-SFT, while 
    text-covered pairs are reserved for counterfactual augmentation. Stage 2 
    in the middle shows visual grounding with a red dress example, where 
    visual attributes are removed from the product text to break textual 
    shortcuts. Stage 3 on the right shows utility optimization via GRPO, 
    where a safety gate combines semantic, commercial, and visual rewards.}
     \label{fig:main}
\end{figure*}

Standard approaches typically maximize the likelihood $\mathcal{L}_{MLE} = \log \pi_\theta(Q \mid \mathcal{T}, \mathcal{I})$ over raw Query-Item click logs. However, we argue that naive optimization of MLLMs on noisy industrial logs leads to \textit{objective misalignment}, where the model learns spurious correlations rather than true semantic expansion. To address this, SAM-D2Q adopts a \textbf{Three-Stage Curriculum Learning} pipeline (as illustrated in Figure~\ref{fig:main}), which progressively evolves the MLLM from learning basic semantic associations (Stage 1) to mastering cross-modal reasoning (Stage 2), and finally aligning with complex business utilities (Stage 3).

\subsection{Stage 1: Bridging the Semantic Gap via Information-Gain Constraints}
\label{sec:stage1}
The primary objective of the first stage is to instill the capability of \textit{Lexical Expansion}. A significant challenge in training generative models on raw e-commerce logs is the "summarization shortcut": models tend to converge to a local optimum where they merely extract subsets of tokens from $\mathcal{T}$ to form $Q$, achieving low training loss but \textbf{ zero retrieval gain}.

To bridge this mismatch, we split query-item click logs according
to whether the query provides information gain over the product text.
For Stage-1 pre-SFT, we retain only samples where the query contains
new terms that are not fully covered by the product text:
\begin{equation}
    \mathcal{D}_{\mathrm{preSFT}} =
    \{ (\mathcal{T}, \mathcal{I}, Q) \in \mathcal{D}
    \mid \mathcal{V}_Q \not\subseteq \mathcal{V}_{\mathcal{T}} \},
\end{equation}
where $\mathcal{V}_Q$ and $\mathcal{V}_{\mathcal{T}}$ denote the set of
unique tokens in the query and product text, respectively.

The remaining text-covered pairs, where
$\mathcal{V}_Q \subseteq \mathcal{V}_{\mathcal{T}}$, are not used for
Stage-1 pre-SFT. Instead, they are reserved as candidates for Stage-2
counterfactual visual augmentation. By training Stage 1 on
$\mathcal{D}_{\mathrm{preSFT}}$, the policy $\pi_\theta$ is discouraged
from mere extraction and is encouraged to learn latent semantic mappings
from product context to informative expansion queries.

\subsection{Stage 2: Hybrid Supervision with Counterfactual Visual Augmentation}
\label{sec:stage2}

The first stage enforces lexical expansion by excluding query-in-text
samples from pre-SFT. However, many text-covered pairs still contain
strong relevance signals, especially when they involve visually
expressed attributes. To further improve cross-modal grounding while
preserving the information-gain property learned in Stage 1, we
construct the Stage-2 training corpus from two complementary sources:
CPV-guided multimodal augmentation data and high-quality human-labeled
data.

\subsubsection{CPV-Guided Counterfactual Data Augmentation.}
We leverage the platform's accumulated Category-Property-Value (CPV)
knowledge base to identify visual attributes of products. In the CPV
schema, attributes such as color, material, pattern, and sleeve style
are defined as visual attributes because their values can be directly
perceived from product images.

Unlike Stage 1, which uses log pairs where the query already provides
information gain over the product text, counterfactual augmentation
starts from text-covered log pairs satisfying
$\mathcal{V}_Q \subseteq \mathcal{V}_T$. These samples usually contain
strong relevance signals, but the target query can be generated by
simply copying terms from the product text. To remove such textual
shortcuts, if the query contains visual attribute values defined in the
item's CPV schema, we mask these values from the product text while
keeping the product image unchanged.

Formally, for a text-covered query-item pair $(T,I,Q)$, we construct a
counterfactual textual input:
\[
T^{cf} = \mathrm{Mask}(T, V_Q \cap V_{\mathrm{visual}}),
\]
where $V_Q$ denotes the set of query tokens, and
$V_{\mathrm{visual}}$ denotes the set of visual attribute values derived
from the CPV schema. We then retain only the augmented samples that
satisfy the information-gain constraint after masking:
\[
\begin{aligned}
\mathcal{D}_{\mathrm{MDA}} = \Big\{ (T^{cf}, I, Q) \ \Big|\  & (T,I,Q)\in\mathcal{D},\; V_Q \subseteq V_T, \\
& T^{cf}=\mathrm{Mask}(T,V_Q\cap V_{\mathrm{visual}}),\; V_Q \nsubseteq V_{T^{cf}} \Big\}.
\end{aligned}
\]

The model is trained to generate the original query from the masked
textual input and the image:
\[
\max_{\theta} \log \pi_{\theta}(Q \mid T^{cf}, I).
\]
In this way, the original query is fully supported by the product text
before masking, but becomes lexically absent after removing visual
attributes. The model must therefore rely on the product image to
recover the missing visual information.

\subsubsection{High-Quality Human-Labeled Data with Information-Gain Filtering.}
In addition to the augmented data, we also incorporate high-quality human-labeled query-item pairs to provide reliable relevance supervision. However, to remain consistent with the objective of Stage 1, we strictly remove redundant samples where the query text is fully covered by the product textual information. Specifically, we retain only samples satisfying the same information-gain constraint:
\[
\mathcal{D}_{\mathrm{HQ}}^{IG}
=
\{(T, I, Q) \in \mathcal{D}_{\mathrm{HQ}}
\mid
V_Q \nsubseteq V_T
\},
\]
where $V_T$ denotes the set of unique tokens in the product textual information. This filtering strategy prevents the model from being retrained on trivial text-extraction cases and preserves its ability to generate new and informative query terms.

Finally, the Stage-2 training corpus is formed by combining the CPV-guided augmented data and the filtered high-quality human-labeled data:
\[
\mathcal{D}_{\mathrm{SFT}}^{(2)}
=
\mathcal{D}_{\mathrm{MDA}}
\cup
\mathcal{D}_{\mathrm{HQ}}^{IG}.
\]
By training on this hybrid corpus, SAM-D2Q simultaneously learns robust cross-modal grounding from counterfactual visual augmentation and reliable semantic relevance from human-labeled data, while avoiding degeneration into simple text copying.

\subsection{Stage 3: Business-Aware Preference Optimization}

The final stage addresses the objective misalignment between SFT
and commercial search. While SFT optimizes token-level likelihood,
e-commerce search requires generated queries to maximize practical
business utility. Since retrieval is based on Boolean term matching,
the value of a generated query mainly comes from its newly
introduced terms. We therefore formulate query generation as a
reinforcement learning problem and optimize the policy with Group
Relative Policy Optimization (GRPO).

\subsubsection{Reward Design with Composite Utility}
\label{subsec:rewards}
The key component of our preference optimization is a composite reward
that jointly captures three desiderata of high-quality expansion queries:
\textit{semantic correctness}, \textit{commercial value}, and
\textit{visual grounding}. For a generated query $q$ and item $i$,
we define its utility as:
\[
U(q,i)=I_{\mathrm{gate}}(q,i)
\left(
\lambda_1 R_{\mathrm{sem}}(q,i)
+\lambda_2 R_{\mathrm{biz}}(q,i)
+\lambda_3 R_{\mathrm{vis}}(q,i)
\right).
\]

Here, $I_{\mathrm{gate}}(q,i)$ is a binary safety gate. It sets the
overall reward to zero if the query fails basic validity checks,
such as low query-item relevance or failure to introduce useful new
terms. This prevents the model from exploiting business signals to
generate popular but irrelevant queries.

\paragraph{1) Semantic Consistency Reward.}
The first reward ensures that the generated query remains faithful
to the target product. We use a Qwen3-VL-based query-item relevance
model trained on human-annotated data to score the semantic
consistency between the generated query and the product's
multimodal context, including both textual metadata and product
images:
\[
R_{\mathrm{sem}}(q,i)=f_{\mathrm{rel}}(q,T_i,I_i).
\]
This reward acts as the semantic foundation of the utility function,
penalizing hallucinated or irrelevant expansions.

\paragraph{2) Commercial Value Reward.}
The second reward measures whe\-ther the newly introduced terms
have real business potential. For each term $t$ under category $c$,
we estimate its value using historical query traffic, measured by
search Page Views (PV), and Conversion Rate (CVR):
\[
E(t,c)=PV_{t,c}\cdot CVR_{t,c}.
\]

To reduce the impact of sparsity for long-tail terms, we apply
Bayesian smoothing to obtain a robust category-level estimate:
\[
\hat{E}_{\mathrm{cate}}(t,c)
=
PV_{t,c}
\cdot
\frac{\alpha \bar{\mu}_{\mathrm{cvr},c}+PAY_{t,c}}
{\alpha+PV_{t,c}},
\]
where $\bar{\mu}_{\mathrm{cvr},c}$ is the category-level average CVR,
$PAY_{t,c}$ is the number of paid conversions, and $\alpha$ controls
the smoothing strength.

To handle term-level cold-start cases, we combine the category-specific
estimate with a global term-level estimate. Let
\[
s_c(t,c)=\log(1+\hat{E}_{\mathrm{cate}}(t,c)), \quad
s_g(t)=\log(1+\hat{E}_{\mathrm{global}}(t)).
\]
The business score of term $t$ in category $c$ is:
\[
r_{\mathrm{biz}}(t,c)=
\begin{cases}
\eta s_c(t,c)+(1-\eta)s_g(t), & PV_{t,c}>0,\\
\kappa s_g(t), & PV_{t,c}=0.
\end{cases}
\]

Let $\Delta(q,T_i)=V_q\setminus V_{T_i}$ denote the new terms
introduced by $q$ beyond the product text. The item-conditioned commercial reward is computed as:
\[
R_{\mathrm{biz}}(q,i)
=
\mathrm{Norm}
\left(
\sum_{t\in \Delta(q,T_i)}
r_{\mathrm{biz}}(t,c_i)
\right),
\]
where $c_i$ is the category of item $i$. This reward explicitly
pushes the model toward generating expansion terms with higher
traffic and conversion potential.

\paragraph{3) Visual Attribute Bonus.}
The third reward encourages the model to generate visually grounded
terms that are absent from the product text but meaningful for the
item category. We build a category-specific visual attribute lexicon
from the platform's CPV knowledge base. For each category $c$, we
collect values of CPV attributes marked as visual attributes, such as
color, material, pattern, neckline, and sleeve style:
\[
\mathcal{V}^{\mathrm{vis}}_c
=
\{v \mid v \text{ is a value of a visual CPV attribute in category } c\}.
\]

The visual reward is assigned when a newly introduced query term
belongs to the visual attribute lexicon of the item's category:
\[
R_{\mathrm{vis}}(q,i)
=
\mathbb{I}
\left[
\Delta(q,T_i)
\cap
\mathcal{V}^{\mathrm{vis}}_{c_i}
\neq
\emptyset
\right].
\]
This reward reinforces the visual grounding ability learned in Stage
2 and encourages visually meaningful expansion rather than arbitrary
new terms.

\subsubsection{Policy Optimization with GRPO}

We optimize the generation policy using GRPO, which avoids the
extra value network required by PPO. For each input $x$, we sample
a group of $G$ candidate queries $\{y_1,\ldots,y_G\}$ from the
current policy and compute their utilities. The advantage of each
sample is normalized within the group:
\[
\hat{A}_j
=
\frac{
U(y_j,x)-\mathrm{mean}(\{U(y_k,x)\}_{k=1}^G)
}{
\mathrm{std}(\{U(y_k,x)\}_{k=1}^G)+\epsilon
}.
\]

The core group-relative objective can be written as:
\[
\mathcal{L}_{\mathrm{GRPO}}(\theta)
=
-\mathbb{E}
\left[
\frac{1}{G}
\sum_{j=1}^{G}
\frac{\pi_{\theta}(y_j|x)}
{\pi_{\mathrm{old}}(y_j|x)}
\hat{A}_j
\right].
\]
Here, $x$ denotes the multimodal product context and
$\pi_{\mathrm{old}}$ denotes the behavior policy used to sample candidate
queries. For clarity, we present the simplified objective; our
implementation follows the standard GRPO recipe with policy-ratio
stabilization.

Through this reward design, the model is not merely optimized to
generate likely queries, but is explicitly aligned with three practical
requirements of industrial e-commerce search: relevance, business
value, and visual grounding.

\section{Experiments}
\label{sec:experiments}

\subsection{Experimental Setup}

\subsubsection{Datasets.}
We curate large-scale training data from AliExpress production logs for
the three-stage training pipeline of \textbf{SAM-D2Q}. In \textit{Stage 1},
we extract query-item click pairs and apply the Information-Gain
constraint (Sec.~\ref{sec:stage1}) on the original product text,
retaining pairs where the query is not fully covered by the product
text. This yields {9 million} valid pairs for coarse semantic alignment.

In \textit{Stage 2}, we construct a hybrid multimodal SFT corpus. First,
we use text-covered log pairs where the original query is fully covered
by the product text as candidates for CPV-guided counterfactual
augmentation. By masking visual attribute values from the product text
and retaining samples that satisfy the Information-Gain constraint after
masking, we obtain {500k} high-quality counterfactual samples
(Sec.~\ref{sec:stage2}). Second, we incorporate high-quality
human-labeled query-item pairs after applying Information-Gain filtering
on the original product text. These two sources jointly strengthen
visual grounding while preserving the model's expansion ability.

In the \textit{Alignment} stage, we train the Query-Item relevance
scoring model with {3.5 million} human-annotated pairs, derive dynamic
commercial term weights from AliExpress traffic logs collected from
August to November 2025, and sample \textbf{80k} items as RL exploration
prompts for GRPO alignment.

\begin{table*}[t]
\caption{Performance on inverted-index retrieval. Qual.(M) denotes
average Total Qual. in millions over the query set.}
\label{tab:retrieval_main}
\small
\centering
\renewcommand{\arraystretch}{1.15}

\begin{tabular*}{\textwidth}{@{\extracolsep{\fill}}lcccccc}
\toprule
\multirow{2}{*}{\textbf{Method}} & \multicolumn{2}{c}{\textbf{Top-300}} & \multicolumn{2}{c}{\textbf{Top-1000}} & \multicolumn{2}{c}{\textbf{Top-3000}} \\
\cmidrule(r){2-3} \cmidrule(lr){4-5} \cmidrule(l){6-7}
 & \textbf{\# Rel.} & \textbf{Qual.(M)} & \textbf{\# Rel.} & \textbf{Qual.(M)} & \textbf{\# Rel.} & \textbf{Qual.(M)} \\
\midrule
Original & 107.0 & 1,382 & 248.3 & 3,062 & 454.5 & 5,366 \\
\midrule
Doc2Query-Text & 103.3 & 1,395 & 276.9 & 3,513 & 558.0 & 6,696 \\
MM-SFT (Stage 1) & 111.7 & 1,498 & 288.3 & 3,631 & 566.9 & 6,799 \\
SAM-D2Q (w/o RL) & 112.4 & 1,501 & 290.8 & 3,671 & 579.2 & 6,947 \\
\textbf{SAM-D2Q (Full)} & \textbf{113.4} & \textbf{1,505} & \textbf{296.1} & \textbf{3,750} & \textbf{594.1} & \textbf{7,139} \\
\bottomrule
\end{tabular*}
\end{table*}

\subsubsection{Training Details.}
SAM-D2Q is trained with a two-phase pipe\-line: supervised fine-tuning
(SFT) followed by GRPO alignment. In SFT, we use a two-stage training
strategy with a learning rate of $2\times10^{-5}$ and train each stage
for one epoch. In GRPO, the model is trained for two epochs with a
learning rate of $1\times10^{-6}$, weight decay of $1\times10^{-6}$,
and a global batch size of 256. For each prompt, we sample 64 responses
with $top\_p=0.99$, $top\_k=300$, and temperature 1.4. The maximum
sequence length is 2048 tokens for prompts and 256 tokens for responses.
All experiments use Qwen3-VL-8B as the backbone and are implemented
with BFloat16 precision under DeepSpeed ZeRO-3.

\subsubsection{Evaluation Protocol.}
To mirror the industrial retrieval environment, we construct an inverted
index containing {8.8 million items}, preserving the real-world category
distribution of the platform. For evaluation, we sample {30,000 user
queries} from live traffic, stratified across head, torso, and tail
segments to cover diverse user intents. The offline evaluation is
conducted on the constructed 8.8M-item index without applying the online
active-product deployment filter.

For each query, candidate items retrieved by different methods are
evaluated by a frozen production Query-Item relevance evaluator trained
on human-annotated query-item pairs. The evaluator takes both product
text and image as input and outputs a relevance score. A retrieved item
is counted as relevant if its score is above a threshold calibrated on a
held-out human-labeled validation set. The evaluator is fixed before all
offline experiments and is not trained on generated queries from the
test set. We use these automatic labels as scalable offline proxies and
further validate the final system through online A/B testing.

\subsubsection{Baselines.}
We use different baselines for offline component analysis and online
production evaluation.

\textit{(1) Offline component baselines.}
The offline experiments focus on neural document expansion variants,
aiming to isolate the contribution of multimodal SFT, counterfactual
visual augmentation, and RL alignment. We use 8B-scale backbones for all
neural variants because document expansion is performed offline, where
generation quality is prioritized over inference latency.
\begin{itemize}
    \item \textbf{Original}: Standard Boolean retrieval using only the
    original product titles. This baseline quantifies the vocabulary
    mismatch problem before document-side expansion.
    \item \textbf{Doc2Query-Text}: A text-only expansion model based on
    \textit{Qwen3-8B}, fine-tuned on query-title pairs, serving as the
    unimodal Doc2Query baseline.
    \item \textbf{MM-SFT (Stage 1)}: A multimodal baseline based on
    \textit{Qwen3-VL-8B}, trained only with Stage-1 SFT data, without
    counterfactual visual augmentation or RL alignment.
    \item \textbf{SAM-D2Q (w/o RL)}: An ablation variant trained with
    Stage-2 counterfactual visual augmentation but without Stage-3
    business-aware RL alignment.
\end{itemize}

\textit{(2) Online production baseline.}
Traditional e-commerce expansion strategies, such as CPV-attribute
expansion, category lexicon expansion, synonym expansion, and query-side
rewriting, are part of the AliExpress production search stack and are
therefore evaluated in the online A/B test rather than in the offline
component table. In the online experiment, the control bucket uses this
existing production stack, while the treatment bucket adds SAM-D2Q as an
additional document-side expansion module on top of the production
baseline.

\subsubsection{Evaluation Metrics.}
We use three groups of metrics. For generation quality, we report
\textit{Relevance}, measured by a frozen Query-Item relevance evaluator
trained on human-annotated data, and \textit{Query-Value}, defined as
the average commercial term score in Sec.~\ref{subsec:rewards}.
Query-Value is closely related to the alignment reward and is therefore
used only as an intrinsic diagnostic metric. For retrieval effectiveness,
we report \textbf{\# Rel. Items} and \textbf{Total Qual.}. A retrieved
item is counted as relevant if its frozen Query-Item relevance score
exceeds a threshold calibrated on held-out human labels. Total Qual. is
computed as:
\[
\mathrm{TotalQual@K}
=
\frac{1}{|\mathcal{Q}|}
\sum_{q\in\mathcal{Q}}
\sum_{i\in \mathrm{TopK}(q)}
\mathbb{I}[\mathrm{rel}(q,i)=1]\cdot s_i ,
\]
where $s_i$ is a fixed item-level business quality score computed before
evaluation from historical efficiency signals such as CTR, CVR, and Pay
Count. We report it in millions as \textbf{Qual.(M)}. Unlike the
term-level commercial reward, Total Qual. is an item-level evaluation
metric and is not directly optimized during training. For online impact,
we report CTR, CVR, Pay Count, and GMV.

\subsection{Offline Evaluation}
\subsubsection{Analysis of Generation Quality}
As presented in Table~\ref{tab:gen_quality}, the text-only baseline suffers from severe semantic drift, achieving only 43.11\% relevance due to the lack of visual grounding. Introducing multimodal signals immediately mitigates this, producing a sharp +27.3 p.p. gain in relevance. While counterfactual augmentation in Stage 2 further refines accuracy
by suppressing textual shortcuts, the most critical observation lies in
the \textit{alignment} phase. Although Stage 3 yields only marginal gains in relevance (+0.4 p.p.), it acts as a decisive \textit{value multiplier}, propelling the Commercial Value score to a peak of \textbf{0.981}. This confirms that SFT mainly improves semantic plausibility, while RL alignment is crucial for optimizing the commercial utility of generated queries.
\begin{table}[b] 
\caption{Intrinsic generation quality evaluation.}
\label{tab:gen_quality}
\centering
\small
\renewcommand{\arraystretch}{1.15}
\begin{tabular*}{\columnwidth}{@{\extracolsep{\fill}}lcc}
\toprule
\textbf{Method} & \textbf{Relevance} & \textbf{Query-Value} \\
\midrule
Doc2Query-Text & 43.11\% & 0.962 \\
MM-SFT (Stage 1) & 70.42\% & 0.958 \\
SAM-D2Q (w/o RL) & 74.57\% & 0.969 \\
\textbf{SAM-D2Q (Full)} & \textbf{74.96\%} & \textbf{0.981} \\
\bottomrule
\end{tabular*}
\end{table}
\subsubsection{End-to-End Retrieval Performance}
The offline retrieval experiment focuses on neural document expansion;
traditional CPV-, category-, and synonym-based expansions are included
in the online production baseline. As shown in
Table~\ref{tab:retrieval_main}, SAM-D2Q consistently improves retrieval
across all cutoffs. At Top-3000, it increases \# Rel. Items from 454.5
to 594.1 (\textbf{+30.7\%}) and Total Qual. from 5,366M to 7,139M
(\textbf{+33.0\%}). Compared with Doc2Query-Text, SAM-D2Q also retrieves
more relevant items with higher aggregate quality, showing the benefit
of multimodal grounding and business-aware alignment.

\subsubsection{Task Adaptability}
We further test whether SAM-D2Q benefits other search modules. As shown
in Table~\ref{tab:adaptability}, adding SAM-D2Q to a production QR
system improves Top-3000 relevance and quality, indicating that query-side
rewriting and document-side expansion are complementary. Concatenating
SAM-D2Q terms to item titles also improves dense retrieval with
Qwen-Emb-4B, showing that multimodal expansion provides useful textual
signals even for embedding-based retrieval.

\begin{table}[b]
\caption{Adaptability on downstream tasks.}
\label{tab:adaptability}
\centering
\small
\renewcommand{\arraystretch}{1.15}
\begin{tabular*}{\columnwidth}{@{\extracolsep{\fill}}lcc}
\toprule
\textbf{Method} & \textbf{\# Rel.} & \textbf{Qual.(M)} \\
\midrule
\multicolumn{3}{l}{\textit{{Task 1: Interaction with Query Rewriting}}} \\
\cmidrule(r){1-3} 
QR + Baseline & 532.4 & 6,391 \\
\textbf{QR + SAM-D2Q} & \textbf{616.8} & \textbf{7,524} \\
\midrule
\multicolumn{3}{l}{\textit{{Task 2: Augmenting Dense Retrieval}}} \\
\cmidrule(r){1-3}
Original (Title) & 581.8 & 5,497 \\
\textbf{+ SAM-D2Q Text} & \textbf{627.5} & \textbf{5,945} \\
\bottomrule
\end{tabular*}
\end{table}

\subsection{Online Deployment and Evaluation}

To validate the real-world business impact of SAM-D2Q, we deployed it
in AliExpress Search and conducted a 21-day A/B test on 4\% of live
traffic. The control bucket used the existing production retrieval and
ranking stack, including CPV-attribute expansion, category lexicon
expansion, synonym expansion, query-side rewriting, and other recall
channels, but without SAM-D2Q document-side expansion. The treatment
bucket kept all other components unchanged and only added the
SAM-D2Q-enhanced document-side expansion index.

SAM-D2Q is deployed through offline indexing and introduces no online
model inference. It is applied to active products with established
purchase history. For each selected product, SAM-D2Q generates Top-100
candidate queries, which are inserted into the inverted index after
relevance and safety filtering. The enhanced index increases index size
by about 35\%, while P99 recall latency increases by only 0.2\%. The
system supports daily incremental updates, and low-quality or invalid
expansions can be removed through the same mechanism. We refresh
commercial term weights and perform GRPO-based alignment updates every
two months. Before indexing, all generated queries pass rule-based
filters for sensitive terms, prohibited terms, risky expressions, and
key-brand protection.

\paragraph{Online A/B Test Results.}
The A/B test was conducted from June 8 to June 28, 2025. Users were randomly
assigned by the platform experimentation system, and all reported numbers
are relative lifts over the production baseline. Pay Count and CVR passed
the platform's standard 95\% confidence checks; other metrics are
reported as directional business indicators.

As shown in Table~\ref{tab:online_ab}, SAM-D2Q improves all core
business metrics on overall search traffic, increasing GMV by
\textbf{+3.38\%}, Pay Count by \textbf{+2.27\%}, UV Value by
\textbf{+2.98\%}, CTR by \textbf{+0.25\%}, and CVR by
\textbf{+0.53\%}. The gains are also consistent on the core five
countries. SAM-D2Q reduces the zero-result PV ratio by \textbf{17\%}
from 2.07\% to 1.72\%, suggesting improved sparse-retrieval coverage.
After full rollout, a 14-day reverse-bucket validation disabled SAM-D2Q
for a small holdout bucket; CTR, CVR, Pay Count, and GMV dropped by
0.75\%, 0.24\%, 1.74\%, and 2.80\%, respectively, further supporting the
causal impact of SAM-D2Q.

\begin{table}[t]
\centering
\caption{Online A/B test results of SAM-D2Q on AliExpress Search
(June 8--28, 4\% live traffic).}
\label{tab:online_ab}
\small
\begin{tabular}{lcccc}
\toprule
Segment & GMV & Pay Count & CTR & CVR \\
\midrule
Overall & +3.38\% &  +2.27\% & +0.25\% & +0.53\% \\
Core-5 Countries & +3.03\% & +1.99\% & +0.23\% & +0.22\% \\
\bottomrule
\end{tabular}
\end{table}

\section{Conclusion}
We present SAM-D2Q, a multimodal document-to-query framework for
industrial e-commerce search. SAM-D2Q bridges vocabulary mismatch
in sparse retrieval by generating semantically relevant, visually
grounded, and commercially valuable expansion queries. It combines
multimodal SFT, counterfactual data synthesis, and business-aware
GRPO alignment with semantic, commercial, and visual rewards.
Extensive offline experiments show consistent improvements in
generation quality and retrieval effectiveness, while online A/B
testing on AliExpress Search further demonstrates significant gains
in core business metrics, including GMV and Pay Count. These results
validate the effectiveness and deployment feasibility of SAM-D2Q in
large-scale industrial search systems.

\section*{GenAI Usage Disclosure}
During the preparation of this manuscript, the authors used
large language model-based tools, including ChatGPT and Claude,
to assist with English language polishing and grammar checking.
The authors reviewed and edited all AI-generated suggestions and
take full responsibility for the content of this publication.

\bibliographystyle{ACM-Reference-Format}
\bibliography{reference}

\end{document}